\documentclass[journal, table]{IEEEtran}

\usepackage{amsmath, amssymb, booktabs, bm, cite, color, lipsum, siunitx, mathdots, multirow, textcomp, verbatim, xcolor, cuted, stfloats}
\usepackage{threeparttable}

\usepackage{longtable}
\usepackage{graphicx}
\graphicspath{{figures/}}

\ifCLASSOPTIONcompsoc
  \usepackage[caption=false,font=normalsize,labelfont=sf,textfont=sf]{subfig}
\else
  \usepackage[caption=false,font=footnotesize]{subfig}
\fi

\usepackage{mdwmath}
\usepackage{eqparbox}
\usepackage{url}
\usepackage{algorithm}
\usepackage{algorithmicx}
\usepackage{algpseudocode}

\usepackage{booktabs} 
\usepackage{multirow} 
\usepackage{lipsum}   
\usepackage{tikz}
\usepackage{siunitx}
\usepackage{tabularx}
\usepackage{makecell}
\usepackage{booktabs}
\usepackage{multirow}

\usetikzlibrary{shapes,arrows,fit,positioning}

\usepackage{etoolbox}
\usepackage{hyperref}
\newtoggle{hl}
\togglefalse{hl}
\iftoggle{hl}{%

}{%

}

\def\BibTeX{{\rm B\kern-.05em{\sc i\kern-.025em b}\kern-.08em
    T\kern-.1667em\lower.7ex\hbox{E}\kern-.125emX}}

\begin{document}
\bstctlcite{IEEEexample:BSTcontrol}
    \title{Respiration Monitoring of Multiple People using Multi-site FMCW SISO Radar Systems}
\author{%
\IEEEauthorblockN{%
Lang Qin\textsuperscript{\#\$}, 
Mandong Zhang\textsuperscript{\$}, 
Wenting Song\textsuperscript{\$},
Zhiqiang Huang\textsuperscript{\#},
Xiaoguang Liu\textsuperscript{\$}\\
}
\IEEEauthorblockA{%
\textsuperscript{\#}The Hong Kong University of Science and Technology (Guangzhou), China\\
\textsuperscript{\$}Southern University of Science and Technology, China\\
 zqhuang@hkust-gz.edu.cn, liuxg@sustech.edu.cn
}
}

\maketitle

\begin{abstract}

Continuous contactless respiration monitoring of co-sleeping subjects faces a dilemma: conventional single-site multiple-input multiple-output (MIMO) radars struggle with limited angular resolution for closely spaced individuals, while distributed radar networks typically require complex hardware synchronization. To address these limitations, this paper proposes non-coherent multi-site single-input-single-output (SISO) radar systems that completely eliminate the need for physical synchronization cables or common reference clocks. The fundamental challenge of ghost target ambiguity in such non-coherent multilateration is resolved through a novel physiological-feature-assisted suppression technique. By exploiting the inherent statistical independence of individual respiratory rhythms, true target locations are robustly distinguished from ghosts via cross-correlation analysis. Experimental validation demonstrates that the proposed system can accurately resolve two subjects spaced less than 20 cm apart, surpassing the resolution limits of traditional compact MIMO arrays, while achieving a respiration rate estimation accuracy of 0.7 bpm root mean square error (RMSE) compared to contact-based ground truth.

\end{abstract}

\renewcommand{\IEEEkeywordsname}{Keywords}


\section{Introduction}
\begin{figure}
    \centering
    \includegraphics[width=3.25in]{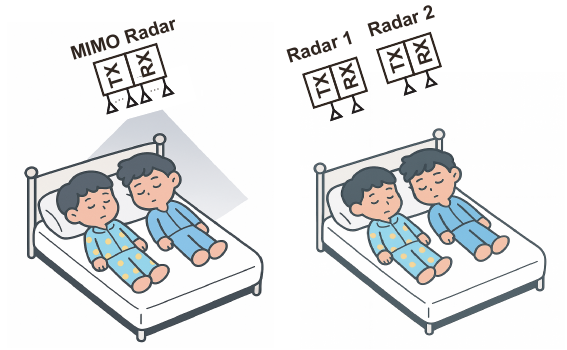}
    \caption{Sleep monitoring scenarios for long-term home vital signs detection. (a) Conventional MIMO radar systems. (b) The proposed multi-site SISO radar systems.}
    \label{fig:main}
\end{figure}

\IEEEPARstart{C}{ontinuous} contactless respiration monitoring using microwave radar has attracted significant attention for sleep assessment~\cite{8755919,li_RobustAccurateFMCW_2024,Zheng_Continuous_2024,Ren_Distributed_2021,Juan_DistributedMIMOCW_2023,10745119,Feng_TMTT2021}. Accurate radar-based respiration monitoring relies heavily on the spatial resolution of the sensing system. In multi-subject scenarios, such as a couple sleeping in the same bed, the thoracoabdominal regions are often closely spaced. Conventional single-site multiple-input multiple-output (MIMO) radars distinguish targets primarily based on angular resolution as shown in Fig.~\ref{fig:main}(a)~\cite{Feng_TMTT2021,zhang2025accurate}. Since the large apertures required for high cross-range resolution are impractical for compact devices, subjects often share the same angular cell. This leads to coupled respiratory signals, complicating motion isolation.

To overcome the inherent physical limitations of single-site systems, multi-site radar networks have emerged as a promising solution~\cite{Ren_Distributed_2021,Juan_DistributedMIMOCW_2023,10745119,qin_Indoor_2024}. By observing targets from widely separated perspectives, these systems significantly enhance spatial diversity. However, state-of-the-art multi-site approaches typically operate in a coherent mode, requiring strict time and phase synchronization via wired connections or shared reference clocks~\cite{Ren_Distributed_2021,Juan_DistributedMIMOCW_2023}. This hardware complexity and the need for precise calibration hinder their widespread adoption in consumer-level sleep monitoring applications.



Non-coherent multi-site architectures offer a flexible and cost-effective alternative by utilizing independent radar units without physical synchronization cables. However, this introduces a fundamental challenge: ghost target ambiguity. In a range-only multilateration system lacking phase coherence, the true pairing of range measurements from different radars is unknown, leading to the generation of false (ghost) targets alongside real ones.

In this work, we propose non-coherent multi-site frequency-modulated continuous-wave (FMCW) SISO radar systems designed for robust multi-person respiration monitoring as shown in Fig.~\ref{fig:main}(b). Unlike existing coherent networks, our approach eliminates the need for hardware synchronization. To resolve the ghost target ambiguity, we introduce a respiration-based cross-correlation technique. This method exploits the physiological uniqueness of respiratory rhythms to correctly associate radar echoes with true target locations. The proposed system is validated experimentally involving different sleeping postures, with accuracy verified against a contact-based respiration belt serving as the ground truth.
Crucially, the experimental validation specifically addresses challenging scenarios where the inter-subject distance is less than $20\,\mathrm{cm}$, demonstrating the system's capability to resolve targets beyond the angular resolution limit of conventional MIMO arrays.

\begin{figure}
    \centering
    \includegraphics[width=3.3in]{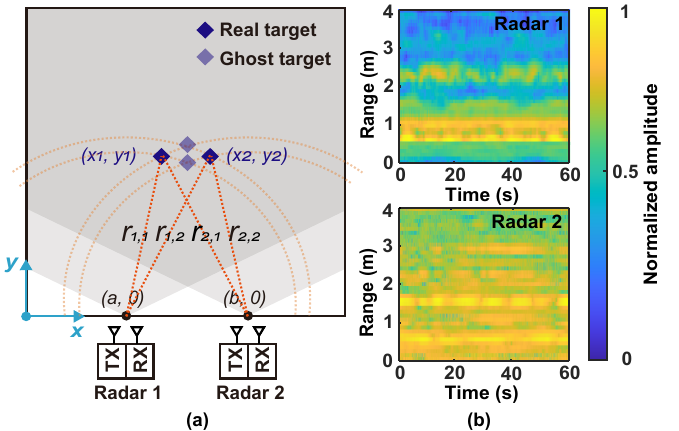}
    \caption{Illustration of the ghost target problem in multi-site multilateration. (a) Ghost target ambiguity schematic. (b) Range-time maps of the two radars showing continuous respiratory motion.
    }
    \label{fig:ghost}
\end{figure}

\begin{figure}
    \centering
    \includegraphics[width=3.3in]{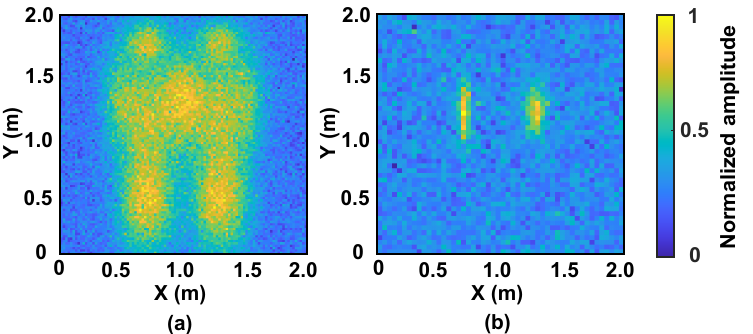}
    \caption{Comparison of multi-person thoracoabdominal positioning results using: (a) Conventional MIMO radar systems. (b) The proposed multi-site radar systems.
    }
    \label{fig:Proposed_method}
\end{figure}
\section{Theory}
After processing the radar data, each radar performs constant false alarm rate (CFAR) detection to extract the presence and the range estimates of the targets. The range measurements are denoted as $r_{i,j}$, where $r_{i,j}$ represents the range estimate of the $j$-th target measured by the $i$-th radar. For dual parallel radar systems, the projection on the target plane is illustrated in Fig.~\ref{fig:ghost}(a). The target's coordinates $(x_{j},y_{j})$ satisfy the following equations
\begin{equation}
\left\{
    \begin{array}{l}
        (x_{j} - a)^2 + {y_{j}}^2 + d^2 = r_{1,j}^2, \\
        (x_{j} - b)^2 + {y_{j}}^2 + d^2 = r_{2,j}^2,
    \end{array}
\right.
\label{eq:values}
\end{equation}

Since the correspondence between a detected range and its originating target is uncertain, two candidate target positions are generated: one corresponds to the real target, and the other to a ghost target. In addition to obtaining the target's coordinate position, the post-detection results can also yield the respiration signal at target's coordinate position. 

\subsection{Signal Model of Respiration Rate Detection}
For two targets sleeping in the bed, the complex form of FMCW radar intermediate frequency (IF) signal within one chirp duration can be expressed as 
\begin{equation}
    S_{b}(t)=\sigma\cdot\exp\left(j\left(\dfrac{4\pi\gamma r_{i,j}(\tau)}{c}t+\dfrac{4\pi f_{c}r_{i,j}(\tau)}{c}+\phi_{1}\right)\right) 
\label{eq:IF signal}
\end{equation}
where $\sigma$ is the amplitude of the IF signal, $\gamma$ is the slope of frequency modulation, $r_{i,j}(\tau)$ is the variation of the distance between the targets and radar, the index $ i \in \{1, 2\} $ denotes each radar unit, $f_{c}$ is the center frequency of the radar, $\tau$ is the slow-time, $c$ is the speed of light, and $\phi_{1}=-4\pi\gamma R^{2}(\tau)/{c^{2}}$ is the residual phase, which can be ignored.

After performing a fast-time domain Range-FFT estimation to identify the accurate range bin where the target is located, the frequency domain expression of the IF signal is as follows:
\begin{equation}
    s_{\text{if}}(f) = \exp \left(j \dfrac{4 \pi f_c r_{i,j}(\tau)}{c}\right) \cdot \operatorname{sinc}\left(T_c\left(f - \dfrac{2 \gamma r_{i,j}(\tau)}{c}\right)\right)
\label{eq:IF signal_2}
\end{equation}
where $T_c$ is the chirp duration and $\mathrm{sinc}(x) = \mathrm{sin}(\pi x)/\pi x$ is the sinc function. 
Letting $f_0 = 2 \gamma r_{i,j}/c$ and extracting the phase of $S_{\text{if}}( f_0)$ along the slow time yields the motion trajectory of the targets, then we can get respiration signal
\begin{equation}
    x_{i,j}(t)  \approx \dfrac{4\pi f_c}{c}  \Delta r_{i,j}(t),
\end{equation}
where $\Delta r_{i,j}(t)$ is the time-varying distance due to respiration. For human targets, faint motion like respiration exists across multiple consecutive range bins as shown in Fig.~\ref{fig:ghost}(b). Consequently, the key challenge is to distinguish real targets from ghost targets and to extract phase information that accurately represents the respiratory motion.
\subsection{Proposed Method}
For each range bin \( m \) in Radar~1 and each range bin \( k \) in Radar~2 exhibiting discernible respiratory activity, 
the cross-correlation function between the corresponding respiratory signals \( x_{1,m}(t) \) and \( x_{2,k}(t) \) is computed as
\begin{equation}
    R_{m,k}(\tau') = \int_{-\infty}^{\infty} x_{1,m}(t)\, x_{2,k}(t + \tau')\, dt,
\label{eq:correlation}
\end{equation}
where \( \tau' \) denotes the time delay. 
For each pair \((m, k)\), the cross-correlation magnitude is analyzed to identify the time lag corresponding to the dominant peak:
\begin{equation}
    \hat{\Delta\tau}_{m,k} = \arg\max_{\tau'} |R_{m,k}(\tau')|.
\label{eq:max}
\end{equation}
If the peak amplitude exceeds $\gamma_{th} = 0.3$ (the $3\sigma$ noise boundary), which rejects random clutter with a $99.7\%$ statistical confidence level, the corresponding bin pair \((m, k)\) is considered to contain a valid contribution from a common target \( j \). All such pairs are then grouped into the set \( C_j \). The corresponding distance $r_{ij}/c$ will then be substituted into~(\ref{eq:values}) to calculate the thoracoabdominal positions of each target. Fig.~\ref{fig:Proposed_method} compares the positioning results of two subjects obtained using MIMO radar systems and the proposed multi-site radar systems. Through the cross-correlation of different radars, we obtain more accurate thoracoabdominal positions of multiple people by using multi-site radar systems. 

To obtain a robust estimate of the target’s respiratory autocorrelation, 
the normalized and delay-compensated correlations are averaged across all valid pairs in \( C_j \):
\begin{equation}
    \hat{r}_{s_j}(\tau) = \frac{1}{|C_j|} 
    \sum_{(m,k)\in C_j} 
    \frac{R_{m,k}(\tau+\hat{\Delta\tau}_{m,k})}
         {R_{m,k}(\hat{\Delta\tau}_{m,k})}.
\label{eq:autocorrelation}
\end{equation}
Finally, applying the Fourier transform to \( \hat{r}_{s_j}(\tau) \) yields the normalized power spectral estimation, which effectively reveals the respiratory rate.

\section{Simulation and Experiments}
\begin{figure*}
    \centering
    \includegraphics[width=6in]{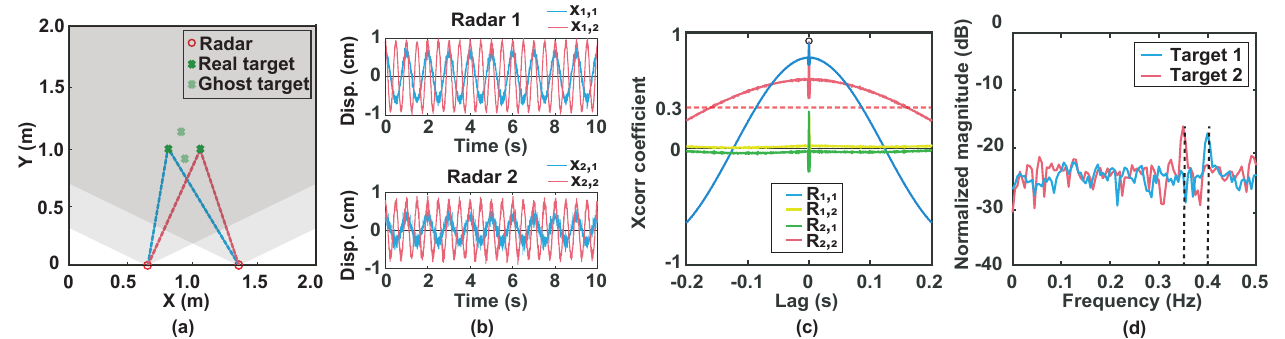}
    \caption{ Simulation results. (a) Geometric configuration of real targets and ghost targets. (b) Extracted respiratory signals from the two radars. (c) Cross-correlation results distinguishing real targets from ghost targets. (d) Normalized respiratory spectra of the two targets.
    }
    \label{fig:correlation}
\end{figure*}

\begin{figure*}
    \centering
    \includegraphics[width=6in]{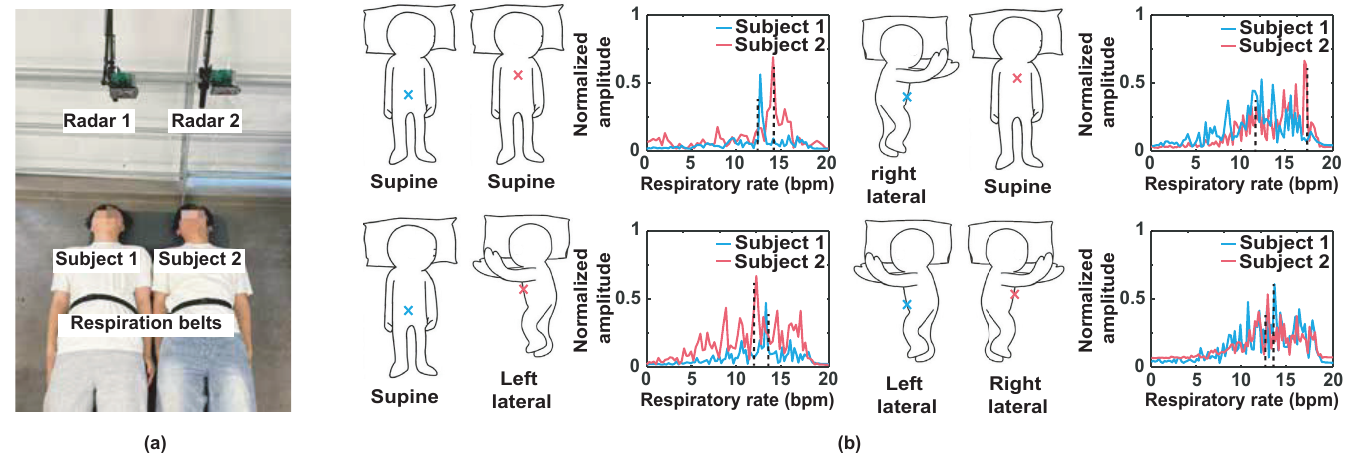}
    \caption{ Experimental results. (a) Photograph of the experimental setup. (b) Normalized respiratory rate spectra for different sleeping postures.
    }
    \label{fig:gesture}
\end{figure*}
The simulation uses dual radar systems with identical parameters for both units. The waveform is configured with a start frequency of \SI{60}{GHz} and a bandwidth of \SI{1500}{MHz}, corresponding to a range resolution of \SI{0.10}{\meter}. The pulse repetition interval (PRI) is set to \SI{0.02}{\milli\second}, with $1024$ pulses integrated per frame. The radars are located at coordinates $(0.50, 0.00)\,\mathrm{m}$ and $(1.50, 0.00)\,\mathrm{m}$. Two targets are positioned at coordinates $(0.80, 1.00)\,\mathrm{m}$ and $(1.10, 1.00)\,\mathrm{m}$. To simulate respiratory motion, the targets are subjected to sinusoidal vibrational motions at frequencies of \SI{0.35}{Hz} and \SI{0.4}{Hz}, each with an amplitude of \SI{1}{cm}. In the simulation, perfect temporal synchronization is assumed to strictly validate the geometric ghost suppression theory. This isolates the impact of phase incoherence from time misalignment errors.

As shown in Fig.~\ref{fig:correlation}(a), the post-detection range from the two targets to the radar are measured as $ r_{11}=1.0 \,\mathrm{m}$, $ r_{12}=1.2 \,\mathrm{m}$, $ r_{21}=1.2 \,\mathrm{m}$ and $ r_{22}=1.1 \,\mathrm{m}$. The candidate target locations can be obtained through~(\ref{eq:values}). These candidate positions are $(0.78,0.96)\,\mathrm{m}$, $(1.12,1.03)\,\mathrm{m}$ and $(0.90,0.92)\,\mathrm{m}$, $(1.00,1.09)\,\mathrm{m}$.
Then the two radars independently monitor the vibration signals. Radar~1 obtains signal \( x_{1,1}(t) \) at distance $ r_{11}$ and \( x_{1,2}(t) \) at distance $ r_{12}$, while Radar~2 obtains signal \( x_{2,1}(t) \) at distance $ r_{21}$ and \( x_{2,2}(t) \) at distance $ r_{22}$. These vibration signals are shown in Fig.~\ref{fig:correlation}(b). The variation of the cross-correlation coefficients with time lag is shown in Fig.~\ref{fig:correlation}(c). The cross-correlation coefficients $R_{1,1}$ and $R_{2,2}$ are above the threshold. This implies that \( x_{1,1}(t) \) and \( x_{2,1}(t) \) originate from the same target, and \( x_{1,2}(t) \) and \( x_{2,2}(t) \) originate from the same target. Finally, normalized power spectral estimation of respiration rate is presented in Fig.~\ref{fig:correlation}(d).

To determine which candidate set corresponds to the real target positions, we analyze the motion along each range. We find that each target’s motion has a one-to-one correspondence with its distance from the radar. For instance, the blue lines represent one target and the red lines another. Based on this relationship, the true target locations are identified as $(0.78, 0.96)\,\mathrm{m}$ and $(1.12, 1.03)\,\mathrm{m}$.

To validate the effectiveness of the proposed method in practical scenarios, a series of experiments was designed. The experiment was conducted in a laboratory environment.
The two radar units were connected to a central host computer and triggered via software commands. Due to the lack of a common reference clock (non-coherent architecture), the recorded data streams exhibited a random initial time offset. Crucially, the proposed cross-correlation method~(\ref{eq:max}) inherently compensates for this temporal misalignment. By searching for the peak lag $\hat{\Delta\tau}_{m,k}$, the system automatically aligns the respiratory signals from the two perspectives, eliminating the need for complex hardware synchronization cables.
Two subjects lay on the bed and the setup is illustrated in Fig.~\ref{fig:gesture}(a). The experimental setup used two Texas Instruments AWR6843ISK millimeter wave radar sensors. To facilitate advanced signal processing, each radar unit was interfaced with a DCA1000EVM real-time data capture adapter to stream raw data to the host computer via Ethernet. Only a single transmit and a single receive antenna were used. The waveform parameters and coordinates of the two radars are the same as in the simulation described previously.

The radars were positioned \SI{100}{cm} above the bed surface, horizontally $d = 75\,\mathrm{cm}$ away from the thoracoabdominal region of the subjects. To validate the measurement accuracy, a contact-based respiration belt was used to record the ground truth respiratory signal simultaneously. Subjects were asked to sleep in three different postures on the bed: supine, left lateral and right lateral. We carefully designed four common scenarios for two-person sleep postures. The inter-subject distance at the torso level was maintained at less than $ 20\,\mathrm{cm}$ across all tested sleeping postures. To evaluate the system's long-term stability and robustness against natural physiological variations, continuous monitoring was conducted for a duration of \SI{30}{min} for each sleeping posture. The dataset comprises over \SI{1800}{s} of radar and ground truth data per scenario.
\begin{table}[h]
    \centering
    \caption{Comparison of Different Sleeping postures}
    \rowcolors{4}{gray!20}{white}
    \begin{tabular}{@{}lcccc@{}}
        \toprule
        \midrule
        & \multicolumn{2}{c}{Subject 1} & \multicolumn{2}{c}{Subject 2} \\
        \cmidrule(lr){2-3} \cmidrule(lr){4-5}
        & Ref.(bpm) & Meas.(bpm) & Ref.(bpm) & Meas.(bpm) \\
        \midrule
        s-s & 12.2 & \textcolor{blue}{12.7} & 15.0 & \textcolor{red}{15.8} \\
        r-s & 13.0 & \textcolor{blue}{13.9} & 17.1 & \textcolor{red}{17.4} \\
        s-l & 14.5 & \textcolor{blue}{13.8} & 13.0 & \textcolor{red}{12.6} \\
        l-r & 14.1 & \textcolor{blue}{13.6} & 12.9 & \textcolor{red}{13.6} \\
        \midrule
        \bottomrule
    \end{tabular}
\begin{tablenotes}
\small
\item s: supine,\; l: left lateral,\; r: right lateral.
\small
\item Averaged over 30-min continuous monitoring.
\end{tablenotes}
\label{tab:postures}
\end{table}

The normalized spectrum estimation of respiration rate of different sleeping postures of two people are shown in Fig.~\ref{fig:gesture}(b). Table~\ref{tab:postures} shows the respiration rate of different sleeping postures comparison between measurements and references. The proposed approach conducts comprehensive evaluations of respiratory rate estimation performance on two participants and achieves a root mean square error (RMSE) of \SI{0.7}{bpm}. This precision supports advanced sleep analysis, such as cardio-respiratory coupling assessment~\cite{Bartsch_cardiorespiratory_2014,Long_Analyzing_2014}. Based on the above analysis, it can be concluded that the proposed approach delivers robust performance across various sleep posture scenarios.

\section{Conclusion}
This paper presents a non-coherent multi-site FMCW SISO radar system capable of multi-person localization and respiration rate estimation for sleep monitoring applications. The proposed approach resolves the ambiguity of multiple candidate targets in non-coherent configurations by using distinct respiration frequencies to identify true torso positions. The cross-correlation of respiratory signals from different range bins is used to associate them with the same subject. Averaging the normalized and delay-compensated correlations across valid pairs yields robust respiration estimation. Experiments with two commercial radars and two subjects in different postures show accurate performance, achieving an RMSE of \SI{0.7}{bpm} across typical two-person sleeping posture scenarios. These results validate the feasibility and accuracy of the proposed systems for non-contact multi-person sleep monitoring.

\section*{Acknowledgment}
This work is supported by the Shenzhen Science and Technology Program under Grant (JCYJ20230807091814030, JCYJ20220818100408018 and 20231115204236001), the National Natural Science Foundation of China under Grant (62471211 and 32371992), and in part by the Guangdong Basic and Applied Basic Research Foundation under Grant (2025A1515011109 and 2024A1515011902).

\bibliographystyle{IEEEtran}
\bibliography{lumped}

\vfill


\end{document}